\documentclass[letterpaper, 10 pt, conference]{ieeeconf}
\IEEEoverridecommandlockouts  % This command is only needed if you want to use the \thanks command

\usepackage{fix-cm}
\usepackage{etex}

\usepackage{dblfloatfix}

\usepackage{nag}

\makeatletter
\@ifpackageloaded{xcolor}{}{%
\usepackage[table,x11names,dvipsnames,svgnames]{xcolor}%
}
\makeatother

\usepackage{colortbl}

\usepackage{graphicx}
\graphicspath{{./figures/}}

\usepackage{wrapfig}

\definecolorset{RGB}{lyft}{}{Red,194,39,36;Sunset,202,53,33;Orange,205,68,20;Amber,200,117,42;Yellow,242,169,52;Citron,186,188,44;Lime,112,159,33;Green,56,139,31;Mint,45,118,56;Teal,52,133,135;Cyan,60,132,202;Blue,55,94,248;Indigo,64,13,247;Purple,115,42,248;Pink,176,25,145;Rose,176,32,75}

\definecolorset{HTML}{h}{}{grapefruit,ED5565;bittersweet,FC6E51;sunflower,FFCE54;grass,A0D468;mint,48CFAD;aqua,4FC1E9;bluejeans,5D9CEC;lavender,AC92EC;pinkrose,EC87C0;lightgray,F5F7FA;mediumgray,CCD1D9;darkgray,656D78}

\definecolorset{RGB}{p}{}{ivory,221,221,221;navy,046,037,133;pine,051,117,056;aqua,093,168,153;azure,148,203,236;sand,220,205,125;peach,194,106,119;carnation,159,074,150;mulberry,126,041,084}

\definecolorset{RGB}{o}{}{black,000,000,000;mint,000,158,115;jeans,000,114,178;azure,086,180,233;banana,240,228,066;orange,230,159,000;ember,213,094,000;amaranth,204,121,167}

\usepackage{cite}

\usepackage{microtype}

\usepackage[american]{babel}

\usepackage{array}
\usepackage{multirow}
\usepackage{booktabs}
\usepackage{makecell} %introduces \thead and \makecell

\ifcsname labelindent\endcsname
\let\labelindent\relax
\fi
\usepackage[inline]{enumitem}

\usepackage{subfig}

\newenvironment{lenumerate}[2][]
{\begin{enumerate}[label=(#2\arabic*),leftmargin=0.2in,itemindent=0.15in,#1]}
{\end{enumerate}}

\setlist*[enumerate,1]{label={\itshape\arabic*)}}

\makeatletter
\newcommand{\paragraphswithstop}{%
\let\copyparagraph\paragraph%
\renewcommand\paragraph[1]{\copyparagraph{##1.}}%
}
\makeatother

\usepackage[framemethod=tikz]{mdframed}

\makeatletter
\def\namedlabel#1#2{\begingroup
  #2%
  \def\@currentlabel{#2}%
  \phantomsection\label{#1}\endgroup
}
\makeatother

\makeatletter
\def\namedlabelphantom#1#2{\begingroup
  \def\@currentlabel{#2}%
  \phantomsection\label{#1}\endgroup
}
\makeatother

\newcommand{\parunskip}{\bgroup\unskip\parfillskip=0pt \par\egroup}

\input{preamble/math}
\newcommand{\real}[1]{\mathbb{R}^{#1}{}}

\newcommand{\bmat}[1]{\begin{bmatrix}#1\end{bmatrix}}

\newcommand{\smallbmat}[1]{\left[\begin{smallmatrix}#1\end{smallmatrix}\right]}

\newcommand{\transpose}{^\top}

\newcommand{\inverse}{^{-1}}

\newcommand{\defeq}{\doteq}

\DeclarePairedDelimiter{\norm}{\lVert}{\rVert}

\newcommand{\vct}[1]{\mathbf{#1}}

\DeclareMathOperator{\stack}{stack}

\newcommand{\compose}{\circ}
\newcommand{\kron}{\otimes}

\newcommand{\subjectto}{\textrm{subject to}\;}

\providecommand{\vs}{\vct{s}}

\providecommand{\vu}{\vct{u}}

\providecommand{\vx}{\vct{x}}

\providecommand{\vpsi}{\bm{\psi}}

\providecommand{\vPsi}{\bm{\Psi}}

\providecommand{\cF}{\mathcal{F}}

\providecommand{\cK}{\mathcal{K}}

\usepackage{units}

  \newcommand{\newcolorlabel}[2]{%
  \expandafter\newcommand\csname #1\endcsname[1]{%
    \tikz[baseline]{\node[text=white,fill=#2,anchor=base,text height=1.3ex,text depth=0.1ex,font=\sffamily\bfseries]{##1}}}%
}

\newcommand{\newcommenter}[2]{%
  \expandafter\newcommand\csname #1\endcsname[1]{%
    \fcolorbox{#2}{#2}{\color{white}\textsf{\textbf{#1}}}
    {\color{#2}##1}}%
  \expandafter\newcommand\csname at#1\endcsname{%
    \fcolorbox{#2}{#2}{\color{white}\textsf{\textbf{@#1}}}
    {\color{#2}}}%
  \expandafter\newcommand\csname #1cite\endcsname[1]{%
    \csname #1\endcsname {[##1]}
  }%
  \expandafter\newcommand\csname #1ref\endcsname[1]{%
    \csname #1\endcsname {$\blacktriangleright$##1}
  }%
  \expandafter\newcommand\csname #1hl\endcsname[2]{%
    \colorbox{#2}{\color{white}\textsf{\textbf{#1}}}\sethlcolor{Azure2}\hl{##2}~%
    \expandafter\ifx\csname commentarrow\endcsname\relax$\leftarrow$\else \commentarrow[#2]\fi~%
    {\color{#2}##1}}%
  \expandafter\newcommand\csname #1st\endcsname[2]{%
    \colorbox{#2}{\color{white}\textsf{\textbf{#1}}}\sout{##2}~%
    \expandafter\ifx\csname commentarrow\endcsname\relax$\leftarrow$\else \commentarrow[#2]\fi~%
    {\color{#2}##1}}%
}
\newcommenter{TODO}{DodgerBlue1}
\newcommenter{rtron}{OliveDrab2}

\usepackage{comment}

\usepackage{pdfcomment}

\usepackage{soul}

\usepackage[normalem]{ulem}

\usepackage{csquotes}

\usepackage{suffix}

\usepackage{environ}

\makeatletter
\newsavebox{\boxifnotempty}
\newcommand{\displayifnotempty}[3]{\sbox\boxifnotempty{#2}\setbox0=\hbox{\usebox{\boxifnotempty}\unskip}%
  \ifdim\wd0=0pt
  \else
  #1\usebox{\boxifnotempty}#3%
  \fi%
}

\newcommand{\ifempty}[2]{\setbox0=\hbox{#1\unskip}%
  \ifdim\wd0=0pt%
  #2%
  \fi%
}

\newcommand{\ifnotempty}[2]{\setbox0=\hbox{#1\unskip}%
  \ifdim\wd0>0pt%
  #2%
  \fi%
}
\makeatother

\newcommand{\switchifempty}[3]{\sbox\boxifnotempty{#1}\setbox0=\hbox{\usebox{\boxifnotempty}\unskip}%
  \ifdim\wd0=0pt{}%
  #2%
  \else{}%
  #3%
  \usebox{\boxifnotempty}%
  \fi%
}

\makeatletter
\@ifundefined{chapter}{\usepackage{algorithm}}{\usepackage[chapter]{algorithm}}
\makeatother
\usepackage{algorithmicx}
\usepackage{algpseudocode}
\makeatother%

\usepackage{scrlfile}

\makeatletter
\newcommand*\newstoreddef[1]{
  \BeforeClosingMainAux{%
    \immediate\write\@auxout{%
      \string\restoredef{#1}{\csname #1\endcsname}%
    }%
  }%
}
\newcommand*{\restoredef}[2]{% used at the aux file
  \expandafter\gdef\csname stored@#1\endcsname{#2}%
}
\newcommand*{\storeddef}[1]{
  \@ifundefined{stored@#1}{0}{\csname stored@#1\endcsname}%
}
\makeatother

\usepackage{pageslts}
\NewEnviron{tee}{\BODY\typeout{Marker Tee [start] ^^J \BODY ^^JMaker Tee [end]}}

\usepackage{cleveref}
\usepackage{nameref}
\makeatletter
\@ifundefined{corollary}{}{
  \crefname{corollary}{corollary}{corollary}
  \Crefname{corollary}{Corollary}{Corollaries}
  \crefname{example}{example}{examples}
  \Crefname{example}{Example}{Examples}
  \crefname{remark}{remark}{remarks}
  \Crefname{remark}{Remark}{Remarks}
  \crefname{property}{property}{properties}
  \Crefname{property}{Property}{Properties}
  \crefname{assumption}{assumption}{assumptions}
  \Crefname{assumption}{Assumption}{Assumptions}
  \crefname{problem}{problem}{problems}
  \Crefname{problem}{Problem}{Problems}
  \crefname{fact}{fact}{facts}
  \Crefname{fact}{Fact}{Facts}
  \crefname{lemma}{lemma}{lemmas}
  \Crefname{lemma}{Lemma}{Lemmas}
  \crefname{proposition}{proposition}{propositions}
  \Crefname{proposition}{Proposition}{Propositions}
}

\usepackage{tikz}
\usetikzlibrary{calc}
\usetikzlibrary{matrix}
\usetikzlibrary{chains,scopes}
\usetikzlibrary{shapes.geometric}
\usetikzlibrary{arrows.meta}
\usetikzlibrary{decorations.markings}
\usetikzlibrary{decorations.pathreplacing}
\usetikzlibrary{backgrounds}

\tikzset{
  dim above/.style={to path={\pgfextra{
        \pgfinterruptpath
        \draw[>=latex,|->|] let
        \p1=($(\tikztostart)!1.5em!90:(\tikztotarget)$),
        \p2=($(\tikztotarget)!1.5em!-90:(\tikztostart)$)
        in(\p1) -- (\p2) node[pos=.5,sloped,above]{#1};
        \endpgfinterruptpath
      }
    }
  },
  dim double above/.style={to path={\pgfextra{
        \pgfinterruptpath
        \draw[>=latex,|->|] let
        \p1=($(\tikztostart)!3em!90:(\tikztotarget)$),
        \p2=($(\tikztotarget)!3em!-90:(\tikztostart)$)
        in(\p1) -- (\p2) node[pos=.5,sloped,above]{#1};
        \endpgfinterruptpath
      }
    }
  },
  dim below/.style={to path={\pgfextra{
        \pgfinterruptpath
        \draw[>=latex,|->|] let
        \p1=($(\tikztostart)!-1em!-90:(\tikztotarget)$),
        \p2=($(\tikztotarget)!-1em!90:(\tikztostart)$)
        in (\p1) -- (\p2) node[pos=.5,sloped,below]{#1};
        \endpgfinterruptpath
      }
    }
  },
}

\tikzset{
    right angle quadrant/.code={
        \pgfmathsetmacro\quadranta{{1,1,-1,-1}[#1-1]}     % Arrays for selecting quadrant
        \pgfmathsetmacro\quadrantb{{1,-1,-1,1}[#1-1]}},
    right angle quadrant=1, % Make sure it is set, even if not called explicitly
    right angle length/.code={\def\rightanglelength{#1}},   % Length of symbol
    right angle length=2ex, % Make sure it is set...
    right angle symbol/.style n args={3}{
        insert path={
            let \p0 = ($(#1)!(#3)!(#2)$) in     % Intersection
                let \p1 = ($(\p0)!\quadranta*\rightanglelength!(#3)$), % Point on base line
                \p2 = ($(\p0)!\quadrantb*\rightanglelength!(#2)$) in % Point on perpendicular line
                let \p3 = ($(\p1)+(\p2)-(\p0)$) in  % Corner point of symbol
            (\p1) -- (\p3) -- (\p2)
        }
    }
}

\newcommand{\pgfextractangle}[3]{%
    \pgfmathanglebetweenpoints{\pgfpointanchor{#2}{center}}
                              {\pgfpointanchor{#3}{center}}
    \global\let#1\pgfmathresult
}

\usetikzlibrary{shapes.arrows}
\newcommand{\commentarrow}[1][Azure4]{\tikz[baseline=-3pt]{\node[shape border uses incircle, fill=#1,rotate=180,single arrow, inner sep=1pt, minimum size=6pt, single arrow head extend=2pt]{};}}

\tikzset{ax/.style={-latex,line width=2pt}}

\tikzset{camera/.style={fill=Sienna1,fill opacity=0.5},%
image plane/.style={draw=RoyalBlue3,line width=2pt}}

\newcommenter{asabelhaus}{DarkOrange1}
\newcommenter{taha}{Cyan3}
\newcommenter{ran}{purple}
\newcommenter{todo}{Firebrick1}
\newcommenter{fordrew}{ForestGreen}
\NewDocumentCommand\kcf{}{KCF}
\WithSuffix\NewDocumentCommand\kcf*{}{Koopman Control Factorization}

\newcommand{\Pin}{\vPsi_{\textrm{in}}}
\newcommand{\Pout}{\vPsi_{\textrm{out}}}

\title{\LARGE\bf
  \kcf*: \\Data-Driven Convex Controller Design for a Class of Nonlinear Systems
}

\author{Taha Ondogan$^1$, Ran Jing$^1$, Andrew P. Sabelhaus$^{1,2}$, Roberto Tron$^{1,2}$% <-this % stops a space
  \thanks{This work was supported by the U.S. National Science Foundation under Award 2432394.}% <-this % stops a space
  \thanks{$^{1}$T. Ondogan, R. Jing, A.P. Sabelhaus, and R. Tron are with the Department of Mechanical Engineering,
    Boston University, Boston MA 02215, USA
    {\tt\small \{tondogan, rjing, asabelha, tron\}@bu.edu}}%
  \thanks{$^2$A.P. Sabelhaus and R. Tron are also with the Division of Systems Engineering, Boston University, Boston MA 02215, USA.}%
}

\begin{document}

\maketitle
\thispagestyle{empty}
\pagestyle{empty}

%%%%%%%%%%%%%%%%%%%%%%%%%%%%%%%%%%%%%%%%%%%%%%%%%%%%%%%%%%%%%%%%%%%%%%%%%%%%%%%%

\begin{abstract}
  Although Koopman operators provide a global linearization for autonomous dynamical systems, nonautonomous systems are not globally linear in the inputs. State (or output) feedback controller design therefore remains nonconvex in typical formulations, even with approximations via bilinear control-affine terms. We address this gap by introducing the Koopman Control Factorization, a novel parameterization of control-affine dynamical systems combined with a feedback controller defined as a linear combination of nonlinear measurements. With this choice, the Koopman operator of the closed-loop system is a bilinear combination of the coefficients in two matrices: one representing the system, and the other the controller. We propose a set of sufficient conditions such that the factorization holds. Then, we present an algorithm that calculates the feedback matrix via semi-definite programming, producing a Lyapunov-stable closed-loop system with convex optimization. We evaluate the proposed controllers on two canonical examples of control-affine nonlinear systems (inverted pendulums), and show that our factorization and controller successfully stabilize both under properly-chosen basis functions. This manuscript introduces a broadly generalizable control synthesis method for stabilization of nonlinear systems that is quick-to-compute, verifiably stable, data-driven, and does not rely on approximations.
\end{abstract}

\section{Introduction}
% \paragraph{Review of prior work}

% \asabelhaus{To Do: bold all vectors.}

The well-known limitations of nonlinear model-based control have prompted a recent resurgence in exact linearizations of nonlinear systems, which allow control designers to rely on linear synthesis techniques.
A very general exact linearization is the Koopman operator, which can represent nonlinear systems as linear dynamics in a high-dimensional space of \emph{observables} \cite{koopman1931hamiltonian,brunton2021modern}.
Koopman operators has been widely used for \emph{state prediction}\cite{mondal2023efficient,surana2017koopman}, \emph{estimation}\cite{guo2021koopman, khorshidi2024centroidal}, and \emph{system analysis} \cite{arbabi2017study,arbabi2017ergodic,mezic2013analysis} tasks in the last decades. Its applications span a variety of dynamical systems, such as fluid mechanics\cite{arbabi2017study,mezic_analysis_2013}, robotic systems\cite{khorshidi2024centroidal}, power systems\cite{matavalam_datadriven_2024}, aerospace engineering\cite{hofmann_analytical_2025}, and even time-varying systems \cite{gueho_timevarying_2021,hao_deep_2024}.

More significantly, Koopman-based models have been widely adopted for control applications\cite{bruder2021advantages,wang2021immersion,abrahammodel}.
% By lifting nonlinear dynamics into a higher-dimensional space where they evolve linearly, Koopman models can be readily
Global, exact, data-driven models makes controller design more robust and tractable, allowing Koopman methods to be integrated with classical techniques including the linear quadratic regulatior (LQR) and linear model predictive control (LMPC).
% , and can serve as state estimators\cite{surana2017koopman}.
However, the applicability of these methods remains limited: the linearity achieved with respect to observables does not necessarily extend to the control input, which may hinder compatibility with linear optimization-based controllers in certain scenarios.
Moreover, the computational burden of receding-horizon controllers grows rapidly with the dimension of the lifting space, particularly when nonlinear constraints are included in the optimization problem.

Attempts at resolving this challenge typically focus on improved lifting.
One can substitute the lifting functions with a neural network instead, followed by an encoder-decoder\cite{shi2022deep,hao_deep_2024}. 
Despite the stronger modeling capability, such structures suffer from local minima, over-fitting, and heavy computation load in the training process\cite{nozawa2024monte}, along with interpretability issues.
Most importantly, adjusting the lifting functions does not address bilinearity/nonconvexity issues.

To address these challenges, we propose a general convex control design framework based on semidefinite programming (SDP)\cite{vandenberghe_semidefinite_1996} for Koopman-identified nonlinear systems. This paper integrates modeling and control into a unified framework (similar to ``System Identification for Control'' \cite{geversaa2005identification}, now in the context of Koopman), which offers advantages in terms of model validity and parameter search efficiency. 
Our key insight is to factorize the closed-loop Koopman system to retain the linearity of the control input in the lifted space, enabling the design of a fixed state-feedback control law. 
% \asabelhaus{Suggestions from Roberto: contributions should focus on the offline aspect of the computation}
This control law is computed entirely offline, avoiding the iterative optimization in receding-horizon controllers, which often suffer from convergence under strict real-time requirements.
Since convex programs can be solved efficiently in polynomial time \cite{boyd2004convex}, our approach mitigates the “curse of dimensionality” for Koopman-identified systems.
% typical of receding-horizon methods and enables super-fast online update for real-time control. 
Under the assumption that the Koopman approximation faithfully represents the underlying system dynamics, our method offers a practical approach for control-law synthesis with verifiable closed-loop stability.
This framework can be applied to a wide class of real-world systems, as defined below.

This manuscript offers two major contributions to the state of the art via our insight of the \kcf*{} (\kcf).
First, we introduce assumptions that formalize the class of applicable underlying plants, and prove that our approach yields an operator that is bilinear in the system and controller coefficients; that is, the operator is linear in each argument when the other is fixed.
Second, we provide a complete practical recipe that uses the proposed \kcf{} and convex optimization to implement a linear controller in nonlinear features. 
% our procedures includes: lifting function selection \taha{\atran Shall we rephrase it to push "lifting function selection" out of scope of the paper? We cannot justify why we use so fine tuned double pendulum lifting, so I was claiming it was not a concern of the paper. (Actually it is still a concern indirectly because of \cref{assumption:H} and \cref{assumption:K}) must hold}, 
With an appropriate choice of observables, system identification can be performed using Extended Dynamic Mode Decomposition (EDMD), followed by the synthesis of control gains with Lyapunov stability guarantees via SDP. The validity of the method is demonstrated by stabilizing of two inverted pendulum systems.

% \myparagraph{Paper outline}

The paper is organized with a review of state of the art (\Cref{sec:Preliminaries} and \Cref{sec:koopman_operators}) before introducing our factorization and synthesis method (\Cref{sec:synthesis}).
We then demonstrate how the method can be applied to two example control-affine systems (\Cref{sec:evaluations}).
We conclude in \Cref{sec:conclusion} with the implications of our underlying assumptions and prospects for future work.

\section{Preliminaries and Notation}
\label{sec:Preliminaries}
Before presenting the proposed control framework, we review the key mathematical concepts and notation results that will be used throughout the paper.

We use $[\vx]_i$ and $[A]_{ij}$ to denote, respectively, the $i$-th and $i,j$-th entries of the vector $\vx$ and the matrix $A$. We use the superscript ${}^+$ to denote a quantity at the next time step (e.g., $\vx^+$ denotes the state $x$ at the following time step).

\subsection{Kronecker product}
\begin{definition}
  Given two matrices $A$ and $B$, the Kronecker product $A\kron B$ is defined as
  \begin{equation}\label{eq:kron}
    A\otimes B = \bmat{
      [A]_{11} B & \cdots & [A]_{1n}B \\
      \vdots & \ddots &           \vdots \\
      [A]_{m1} B & \cdots & [A]_{mn} B
    },
  \end{equation}
  \vspace{-1mm}
\end{definition}

\begin{lemma}
  The Kronecker product satisfies the following mixed-product property:
  \begin{equation}\label{eq:kron mixed}
    (AC)\kron(BD)=(A\kron B) (C\kron D)
  \end{equation}
\end{lemma}
\begin{corollary}\label{thm:kronecker vectors}
  Given two vectors $a,b$ and a matrix $M$, the Kronecker product satisfies the following property
  \begin{equation*}\label{eq:kron mixed}
    a\kron Mb=(I\kron M)(a\kron b),
  \end{equation*}
  where $I$ is identity (dimensions compatible with $a$).
\end{corollary}

\begin{lemma}
  Let $a,b,c \in \mathbb{R}^n$. Then the Kronecker product $\otimes$ and the Hadamard product $\odot$ satisfy the following identity:
  \begin{equation}
    (a \odot b) \otimes c \;=\; (a \otimes \mathbf{1}_n) \odot (b \otimes c),
    \label{eq:kron-hadamard}
  \end{equation}
  where $\mathbf{1}_n \in \mathbb{R}^n$ denotes the vector of ones.
\end{lemma}

\subsection{Lyapunov stability for Linear Time-Invariant systems}
% Given a square matrix $P$, we use the notation $P\succ0$ to denote positive-definiteness.

% Stability for LTI systems

\begin{proposition}\label{thm:lyapunov}
  Given a Linear Time-Invariant (LTI) system, where $\vx = \vx(t) \in \mathbb{R}^n$ and $\vx^+ = \vx(t+1)$,
  \begin{equation}
    \label{eq:lti}
    \vx^+=A\vx,
  \end{equation}
  the system is stable if and only if the \emph{Lyapunov equation}
  % \rtron{Mention that $+$ represents the quantity at the next discrete time step.}
  \begin{equation}
    \label{eq:lti lyapunov}
    A\transpose PA-P=-Q
  \end{equation}
  has a positive definite solution $P\succ0$ for any $Q\succ0$, where $V(\vx)=\vx\transpose P \vx$ is the corresponding Lyapunov function.
  % Moreover, the following is a valid Lyapunov function for the system
  % \begin{equation}
    % \label{eq:lti lyapunov function}
    % V(x)=x\transpose P x.
  % \end{equation}
\end{proposition}

\begin{remark}\label{remark:lyapunov simplified}
  If we are interested primarily in enforcing stability in an optimization problem, we can eliminate the variable $Q$ by substituting \eqref{eq:lti lyapunov} with the following Linear Matrix Inequality:
  \begin{equation}
    \label{eq:lti lyapunov simplified}
    A\transpose P A-\lambda P\preceq 0,
  \end{equation}
  where $\lambda\in[0,1]$ is a positive scalar that bounds the convergence rate.
\end{remark}

\subsection{Schur complements}
Given an invertible matrix $A$, one can show the following equivalence:
\begin{equation}\label{eq:schur}
  \bmat{A & B\\B\transpose & C}\succ 0 \iff A\succ0, C-B\transpose A\inverse B\succ 0
\end{equation}

%%%%%%%%%%%%%%%%%%%%%%%%%%%%%%%%%%%%%%%%%%
%%% KOOPMAN OPERATOR
%%%%%%%%%%%%%%%%%%%%%%%%%%%%%%%%%%%%%%%%%%
\section{Review of Koopman Operator Theory}
\label{sec:koopman_operators}
Consider a system with discrete-time dynamics $\vx^+=F(\vx)$, where $F$ is, in general, a nonlinear map. Let $\cF$ be the (infinite-dimensional) space of \emph{observable} functions defined on the state space, $\vpsi_i(\vx)$. 
The Koopman operator $\cK:\cF\to\cF$ maps observable functions $\vpsi_i$ to their composition with the dynamics $F$, i.e., $\cK:\vpsi_i\mapsto \vpsi_i \compose F$ \cite{koopman1931hamiltonian,brunton2021modern}.
Given the properties of function composition, $\cK$ is a linear operator; we can therefore select a basis $\{\vpsi_i\}$ for $\cF$, express the Koopman operator as a matrix $K$, and \emph{lift} the system to the linear dynamics $\vpsi^+=K\vpsi$, where the \emph{vectors of observables} $\vpsi\in\real{d_{\vpsi}}$ contain coefficients for the basis in $\cF$.
% Ideally, the basis functions $\vpsi_i$ should be selected to coincide with the eigenfunctions of $\cK$, and the dimension of $\cF$ can be infinite;
In practice, the basis $\{\vpsi_i\}$ is finite-dimensional and is a design choice (e.g., fixed to a Fourier or polynomial basis \cite{williams2015data,korda2018convergence}, reproducing kernel Hilbert spaces \cite{williams2014kernel,khosravi2023representer}, or learned with Deep Neural Networks \cite{li2017extended,yeung2019learning,azencot2020forecasting,lusch2018deep,budivsic2012applied,yao2023koopman}), and can be interpreted as a finite-dimensional \emph{manifold of state features}.

\subsection{Extended Dynamic Mode Decomposition}\label{sec:EDMD}
Given an input-output training dataset $\{\vx_j,\vx^+_j\}$ and a selected set of lifting functions $\vpsi$, the matrix $K$ can be then estimated via the \emph{Extended Dynamic Mode Decomposition} (EDMD, \cite{williams2015data,korda2018convergence,brunton2021modern}), by building the matrices $\Pin=\bmat{\cdots \vpsi(\vx_j) \cdots}$, $\Pout=\bmat{\cdots \vpsi(\vx^+_j)\cdots}$, and then solving the least squares problem:
\begin{equation}\label{eq:edmd}
  K^* = \min_{K}\norm{\Pin-K\Pout}^2
\end{equation}
% \rtron{Make this about EDMD/move to introduction}
as has been widely successful in experimental applications \cite{brunton2021modern,jiang2022correcting,abraham2017model,folkestad2020episodic,shi2023koopman}.
% Koopman theory has found a large number extensions \cite{brunton2021modern,jiang2022correcting} and applications in experimental robotics \cite{abraham2017model,folkestad2020episodic} including soft robotics \cite{shi2023koopman}.

\subsection{Imposing stability of Koopman operators}
There exists methods that impose Schur~\cite{halikias2012strong,boots2007constraint,huang2016learning,gillis2020note,gillis2019approximating} stability (or Hurwitz stability \cite{gantmacher2005applications}, in the continuous-time case) of the Koopman model, but only under specific parametrizations of the system matrix $K$ \cite{mamakoukas2020memory,bevanda2022diffeomorphically,fan2022learning}.
These techniques apply to autonomous systems, where the stability of $\vx^+ = K\vx$ is a system identification question as the plant has no inputs.
This manuscript treats the control synthesis problem instead, where stability of the closed-loop system is the concern.

% \rtron{Also, the goal of this paper is not to identify an already-stable plant, but to synthesize a controller that stabilizes it.}

\subsection{Input-affine systems}
There have been extensions of Koopman theory to systems
\begin{equation}
  \label{eq:original system}
  \vx^+=F(\vx,\vu)
\end{equation}
where $\vu$ is a vector of inputs \cite{huang2018feedback,abraham2017model,abraham2019active,li2019learning,bruder2021advantages}. A straightforward method is to make the observables $\vpsi_i$ a function of both $\vx$ and $\vu$. The resulting model, however, will be nonlinear in $\vu$, thus lacking any substantial benefit for estimation and control. Some work \cite{abraham2017model,abraham2019active} assumes an input-affine (aka, control-affine) form $\tilde{F}(\vx)+B \vu$, which is effectively a local linearization via the fixed $B$ matrix. 
% which is not applicable to complicated physical systems, but results in simple linear feedback controllers.
% The situation can be improved by making assumptions on $F(x,\vu)$, resulting in different special cases that facilitate control; for instance, assuming the form , results in the lifted dynamics $\vpsi^+=K\vpsi+Bu$, for which standard tools from Linear Time Invariant (LTI) systems can be applied .
Alternatively \cite{huang2018feedback, bruder2021advantages}, one could assume observables of the form ${\vpsi}_i(\vx)\kron\vu$. These can exactly model systems that are input-affine (which include most mechanical systems), and result in Koopman models that are \emph{bilinear} in $\vpsi, \vu$:
\begin{equation}\label{eq:koopman affine control}
  \vpsi_x^+=\underbrace{\bmat{K_{xx}& K_{xu}}}_{K_x}\bmat{\vpsi_x\\\vpsi_x\kron \vu},
\end{equation}

where $K_x\in \real{d_{\vpsi}\times (d_{\vpsi}+d_{\vpsi}d_{u})}$, $K_{xx}\in\real{d_{\vpsi}\times d_{\vpsi}}$, and $K_{xu}\in \real{d_{\vpsi}\times (d_{\vpsi}d_{u})}$.

\begin{lemma}\label{thm:bilinear}
 (Adapted from Bruder et al. \cite{bruder2021advantages}.) The bilinear Koopman model \eqref{eq:koopman affine control} provides an input-affine approximation of the form
  \begin{equation}\label{eq:koopman affine model}
    \vpsi_x^+ = A\vpsi_x + B(\vpsi_x)\vu,
  \end{equation}
  that can approximate any input-affine plant
  \begin{equation} \label{eq:plant affine}
    \vx^+ = f(\vx) + g(\vx)\vu.
  \end{equation}
\end{lemma}
\begin{proof}
  For the first part of the claim, we first define a partition of $K_{xu}\in d_{\vpsi_x}\times (d_ud_{\vpsi_x})$  into $d_{\vpsi_x}$ blocks:
  \begin{align}\label{eq:blocks Kxu}
    K_{xu}=\bmat{K_{xu}^{(1)} & \cdots & K_{xu}^{(d_{\vpsi_x})}}.
  \end{align}
  Using the definition in \eqref{eq:kron}, we can then partition the Kronecker product in \eqref{eq:koopman affine control} into a block matrix summations, leading to the equivalent expression
  \begin{equation}
    \label{eq:koopman affine blocks}
    \vpsi_x^+ = K_{xx}\vpsi_x + \big(\sum_{i=1}^{d_{\vpsi_x}} [\vpsi_x]_i K^{(i)}_{xu}\big)u;
  \end{equation}
  which is in the input-affine form of \eqref{eq:koopman affine model}. We refer readers to \cite{bruder2021advantages} for the second claim (generality).
\end{proof}

The bilinear model \eqref{eq:koopman affine control} can still be learned using EDMD \eqref{eq:edmd} with appropriate choices of $\Pin$ and $\Pout$, but control synthesis remains a nonlinear problem requiring e.g. model-predictive control \cite{bruder2020data}.

% Overall, the prior work above only considers \emph{state feedback} (i.e., the entire state of the system is available); in practice, novel systems such as soft or biohybrid robots are infinite-dimensional, and it is necessary to pick (through design) subsets of the state that are sufficient for \emph{output~feedback}.

%%%%%%%%%%%%%%%%%%%%%%%%%%%%%%%%%%%%%%%%%%
%%% KOOPMAN CONTROL FACTORIZATION
%%%%%%%%%%%%%%%%%%%%%%%%%%%%%%%%%%%%%%%%%%
\section{Control Synthesis for Bilinear Koopman Models}\label{sec:synthesis}

Whereas previous work aims to first model a system using a Koopman operator (\Cref{sec:koopman_operators}), and then find a nonlinear controller, we propose a different perspective.
We start with the bilinear Koopman model \eqref{eq:koopman affine control}, and define a parametrized family of \emph{feedback controllers} that takes a \emph{linear} combination of \emph{non-linear} features of the state:
\begin{equation}\label{eq:u Ku}
  \vu=K_u\vpsi_u(\vx),
\end{equation}
where %$y$ is a vector of \emph{measurements} (determined by the hardware of the system),
$\vpsi_u\in\real{d_{\vpsi_u}}$ is a vector of user-defined nonlinear features extracted from the state $\vx$, and $K_u\in\real{d_u\times d_{\vpsi_u}}$ is a matrix containing controller parameters. Our goal is to find the coefficients that stabilize the system around the origin. Note that the control \eqref{eq:u Ku} is linear in $\vpsi_u$, but produces a control field that can still be nonlinear, due to the nonlinearity of the features $\vpsi_u$.

Our key insight is to model not the original control system \eqref{eq:original system}, but the \emph{closed-loop} system $\tilde{F}$, parametrized by the coefficients~$K_u$:
\begin{equation}\label{eq:closed loop}
  \vx^+=F\left(\vx, K_u\vpsi_u(\vx)\right)=\tilde{F}_{K_u}(\vx).
\end{equation}
Although the model \eqref{eq:closed loop} enjoys the asymptotic infinite-dimensional guarantees of Koopman operators for autonomous systems (\Cref{sec:koopman_operators}), it does not facilitate the synthesis of the controller $K_u$ due to the nonlinearity of $\vpsi_u$. The following subsection shows that under some assumptions, \cref{eq:closed loop} reduces to an operator bilinear in the system and control parameters.

\subsection{The Koopman Control Factorization}
\label{sec:kcf}

Our first assumption is on the \emph{compatibility} between the observables $\vpsi_x$ and the features $\vpsi_u$:%\footnote{This formulation facilitates our exposition; its applicability is similar and can be expanded with the notion of \emph{controllability}~\cite{hespanha2018linear}.}
\begin{assumption}\label{assumption:H}
  There exists a \emph{selection matrix} $S\in\real{d_S\times d_{\vpsi_x}}$ and a \emph{measurement matrix} $H\in\real{(d_Sd_{\vpsi_u})\times d_{\vpsi_x}}$ such that
  \begin{equation}\label{eq:assumption}
    \bigl(S\vpsi_x(\vx)\bigr)\otimes \vpsi_u(\vx) = H\vpsi_x(\vx).
  \end{equation}

  % , and \item\label{it:psi constant} One of the observables is constant: $\vpsi_i(x)=1$.
\end{assumption}
% \taha{There was a discussion about small subspace angle.}
% \taha{Should we use $\|S\vpsi_x \kron \vpsi_u - H\vpsi_x\| \leq \epsilon$ instead?}
% \rtron{We should introduce $\epsilon$ if we use it, i.e., if we take it into account in the control synthesis. We will deal with this in a future paper.}
Intuitively, the assumption states that the observables $\vpsi_x$ are \emph{rich enough} to express the product of state features $\vpsi_u$ with the observables themselves; this leads to constraints on the relative \emph{order} of the two sets of functions, and the necessity of the selection matrix, as exemplified in the following.
  \Cref{eq:assumption} is an equation in $\vpsi_x$ (since it appears on both sides of the equality). The matrix $S$ is necessary for this equation to have a solution, as illustrated by the following.
\begin{example}
  \label{example:feature-selection}
 As an example, assume a simple case were $\vx=x\in \mathbb{R}$, $\vpsi_x=\vpsi_u=x$ and $S=1$. Then the LHS of \eqref{eq:assumption} reduces to $x^2$, while the RHS contains only $x$; as a result, there exists no $H$ that can match the order of the terms on the two sides. In this case, instead, we would need to use $\vpsi_x=\smallbmat{x\\x^2}$, $\vpsi_u=x$, $S=\bmat{1&0}$.
\end{example}

\begin{assumption}\label{assumption:K}
  The plant admits a bilinear Koopman model, similar to \eqref{eq:koopman affine control}, of the form  \begin{equation}\label{eq:assumption_koopman_bilinear}
    \vpsi_x^+=K_x\bmat{\vpsi_x\\S\vpsi_x\kron u},
  \end{equation}
  where the selection matrix $S$ is the same in \Cref{assumption:H}.
\end{assumption}
Intuitively, this assumption, when combined with \Cref{thm:bilinear}, implies that the plant is input-affine with control fields $g(\vx)$ that can be expanded with a subset of the observable functions.
This assumption is not restrictive: if \Cref{assumption:K} is violated because $S$ eliminates too many terms from the bilinear control input, observable functions can be redesigned according to \ref{eq:assumption}, as in the following.
\addtocounter{example}{-1}
\begin{example}[Continued]
   Continuing the example, one can choose a richer set of features for $\vpsi_x$, e.g., $\vpsi_x=\smallbmat{x\\x^2\\x^3}$, $\vpsi_u=x$, $S=\bmat{1&0&0\\0&1&0}$, for a scalar system state $x$.
\end{example}

Substituting control law \eqref{eq:u Ku} into \eqref{eq:assumption_koopman_bilinear}, and with some algebraic manipulations, we arrive at our proposed \kcf*.
\begin{theorem}[\kcf*]
  Under \Cref{assumption:H,assumption:K}, the closed-loop dynamics $\tilde{F}$ admits a Koopman operator representation $\vpsi^+=\tilde{K}\vpsi$ where the matrix $\tilde{K}$ is bilinear in the \emph{system matrix} $K_x=\bmat{K_{xx}&K_{xu}}$, and the \emph{control parameters} $K_u$:
  \begin{equation}\label{eq:K tilde}
    \tilde{K}=\bigl(K_{xx}+K_{xu}(I\kron K_u) H \bigr).
  \end{equation}
\end{theorem}
\begin{proof}
  We first substitute the control law \eqref{eq:u Ku} into \eqref{eq:assumption_koopman_bilinear}, and use the properties of the Kronecker product (\Cref{thm:kronecker vectors}) together with \Cref{assumption:H} to simplify the resulting expression:
  \begin{equation}
    \begin{aligned}\label{eq:kxu}
      \vpsi_x^+&=K_x\bmat{\vpsi_x\\S\vpsi_x\kron u}\\
               &=K_x\bmat{\vpsi_x\\S\vpsi_x\kron K_u\vpsi_u(\vx)}\\
               &=K_x\bmat{\vpsi_x\\(I\kron K_u)(S\vpsi_x\kron \vpsi_u)}\\ %+ Q (\kff\kron I)
               &=K_x\bmat{\vpsi_x\\(I\kron K_u)H\vpsi_x}\\ %+ Q (\kff\kron I)
               &=K_x\bmat{I\\(I\kron K_u)H}\vpsi_x\\ %+ Q (\kff\kron I)
               &=\bigl(K_{xx}+K_{xu}(I\kron K_u) H \bigr)\vpsi_x.
    \end{aligned}
  \end{equation}
  The claim follows from the last equality.
\end{proof}

Since the closed-loop dynamics \eqref{eq:u Ku} are an autonomous system, its Koopman representation becomes exact if we allow the basis of observables $\vpsi$ to be infinite-dimensional. Similarly, \Cref{assumption:H,assumption:K} provide the precise conditions under which the \kcf{} is exact. In practice, as is customary, the Koopman operator is approximated with a large but finite-dimensional set of observables; similarly, the conditions of \Cref{assumption:H,assumption:K} might hold only approximately. 

However, with the next section and in the simulation results in \Cref{sec:evaluations}, we show that these approximations can still be used to first identify a model for the system using random inputs, and then synthesize linear-in-the-parameters controllers for nonlinear systems via convex optimization.

% This expression represents a Koopman model of the , and hence falls in the domain of Koopman theory; as a consequence the closed-loop system can be approximated to arbitrary precision by choosing a sufficiently large observable basis $\vpsi_x$. In the reminder of this section, we discuss how this property can be used for system identification and control.

% \begin{remark}
%   \rtron{Is this remark used? If so, it might be better to move it to where it is used.}
%   \taha{Can we somehow explain the matrix partitioning \eqref{eq:blocks} in preliminaries? $K_{xu}$ blocking is used in \eqref{eq:koopman affine blocks} and $H$ blocking is used for SYSID part.}
%   To clarify the structure of $\tilde{K}$, we split $K_{xu}\in d_{\vpsi_x}\times (d_Sd_{\vpsi_x})$ and $H\in (d_Sd_{\vpsi_u})\times d_{\vpsi_x}$ into $d_S$ blocks:

%   \begin{align}\label{eq:blocks}
%     K_{xu}=\bmat{K_{xu}^{(1)} & \cdots & K_{xu}^{(d_S)}},&&H=\bmat{H^{(1)} \\ \vdots \\ H^{(d_S)}};
%   \end{align}
%   then, we have that
%   \begin{equation}\label{eq:ktilde}
%     \tilde{K}=K_{xx}+\sum_{i}^{d_S}K_{xu}^{(i)}K_u H^{(i)}.
%   \end{equation}
% \end{remark}

\subsection{System Identification via Motor Babbling}
\label{sec:sysID_babbling}
Our approach requires collecting data to identify two components: the system matrix $K_x$, and the pair of selection matrix and measurement matrix $(S,H)$ per \Cref{assumption:H}.

\subsubsection{Trajectory generation}
\label{sec:trajectory generation}
To generate the data we propose to use \emph{motor babbling}, i.e., build a dataset of forced responses of the system with random controllers $K_u$. In particular, we generate trajectories of $T$ time steps using sets $\{K_u^{(i)}\}_i$ of random controller coefficients, and sets $\{x_{0j}\}_j$ of initial states. We redefine the data matrices of \Cref{sec:EDMD} to $\Pin=\bmat{\cdots \Pin^{(i,j)} \cdots}$, $\Pout=\bmat{\cdots \Pout^{(i,j)} \cdots}$, where $\Pin^{(i,j)},\Pout^{(i,j)}$ contain data from the trajectories generated by pairing the $i$-th controller with the $j$-th initial conditions; the specific contents depend on which part of the procedure below the least squares optimization \eqref{eq:edmd} is applied to.

%include Kronecker product terms as . We then use  to identify the controlled systems of the form \eqref{eq:koopman affine control}. Additionally, a

\subsubsection{Identification of Selection and Measurement Matrices}
\label{sec:H_sysID}
% \begin{remark}
%   When the selection matrix $S$ is identity, the matrix $H$,  can be found via least squares, solving a problem similar to \eqref{eq:edmd}. \label{remark:selection_matrix}
% \end{remark}
While we do not have a general approach for jointly finding the pair of selection and measurement matrices $S,H$ from \eqref{eq:assumption}, the following proposition helps to decouple the solution for the different blocks, and enables the use of linear least squares.

\begin{proposition}\label{proposition:selection_matrix}
  Assume that the selection matrix $S$ in \Cref{assumption:H} is binary with each row containing exactly one non-zero entry, i.e., $S\in\{0,1\}^{d_S\times d_{\vpsi_x}}$ and $S\vct{1}_{d_{\vpsi_x}}=\vct{1}_{d_S}$; then, there exists a binary vector $\mathbf{s} \in \{0,1\}^{d_{\vpsi_x}}$ and a measurement matrix $\bar{H} \in \mathbb{R}^{d_{\vpsi_x}d_{\vpsi_u} \times d_{\vpsi_x}}$ that satisfy
  \begin{equation}
    (\mathbf{s} \otimes \mathbf{1}_{d_{\vpsi_u}}) \odot (\vpsi_x \otimes \vpsi_u) = \bar{H}\vpsi_x.
    \label{eq:hadamard_kron_identity}
  \end{equation}
\end{proposition}

\begin{proof}
  We define $\mathbf{s}$ to be a masking operator $\mathbf{s} \odot \vpsi_x$ such that $[\mathbf{s}]_i = 1$ if any row of the selection matrix $S$ picks the entry $[\vpsi_x]_i$, and zero otherwise. Furthermore, we partition the matrix $H\in (d_Sd_{\vpsi_u})\times d_{\vpsi_x}$ into $d_S$ blocks:
  \begin{align}\label{eq:blocks H}
    H=\bmat{H^{(1)} \\ \vdots \\ H^{(d_S)}}.
  \end{align}
  These definitions allow us to write
  \begin{equation}\label{eq:s selection}
    (\mathbf{s} \odot \vpsi_x) \kron \vpsi_u = \bar{H}\vpsi_x,
  \end{equation}
  where,
  \begin{equation}
    \bar{H}^{(j)} =
    \begin{cases}
      H^{(i)}, & \text{if } S_{ij} = 1, \\
      \mathbf{0}, & \text{otherwise}.
    \end{cases}
  \end{equation}
  and $\bar{H}^{(j)}\in \mathbb{R}^{d_{\vpsi_u} \times d_{\vpsi_x}}$ is the $j$-th block of the matrix $\bar{H}$ (with a vertical partition similar to \eqref{eq:blocks H}).
  Using the identity \eqref{eq:kron-hadamard}, we have that $(\mathbf{s} \odot \vpsi_x) \kron \vpsi_u = (\mathbf{s} \otimes \mathbf{1}_{d_{\vpsi_u}}) \odot (\vpsi_x \otimes \vpsi_u)$. Combined with \eqref{eq:s selection}, the claim follows.
\end{proof}

\begin{example}
  If the selection matrix $S$ and measurement matrix $H$ are given as:
  \begin{equation}
    S = \begin{bmatrix}
      1 \quad 0 \quad 0 \quad 0 \\
      0 \quad 0 \quad 1 \quad 0 \\
    \end{bmatrix}, \quad H = \begin{bmatrix}
      H^{(1)} \\
      H^{(2)}
    \end{bmatrix},
  \end{equation}
  then the binary vector $\mathbf{s}$ and augmented measurement matrix $\bar{H}$ are
  \begin{equation}
    \mathbf{s} = \begin{bmatrix}
      1 \\ 0 \\ 1 \\ 0
    \end{bmatrix}, \quad \bar{H} =
    \begin{bmatrix}
      H^{(1)} \\ 0 \\ H^{(2)} \\ 0
    \end{bmatrix}.
  \end{equation}
\end{example}
Since $S$ and $H$ need to be estimated by data, we propose to use \Cref{proposition:selection_matrix} and find $H$ and $S$ through $\bar{H}$ and $\mathbf{s}$. We first find a candidate $\bar{H}$ by solving the least squares problem assuming that $\vs=\vct{1}$:
\begin{equation}\label{eq:H_least_squares}
  \min_{\bar{H}} \|\vpsi_x \otimes \vpsi_u - \bar{H}\vpsi_x\|.
\end{equation}
This problem is of the form \eqref{eq:edmd} with data matrices of the form discussed in \Cref{sec:trajectory generation} with $\Pin^{(i,j)}=\bmat{\cdots \vpsi^{(i,j)}_{x,t} \otimes \vpsi^{(i,j)}_{y}(k) \cdots}$ and $\Pout^{(i,j)}=\bmat{\cdots \vpsi^{(i,j)}_{x}(k) \cdots}$, where $k$ is the time step index, $k\in\{0,\ldots,T-1\}$.

To find $\vs$, note that the least squares problem \eqref{eq:H_least_squares} is separable in the blocks of $\bar{H}$, i.e., each block $\bar{H}^{(i)}$ is the minimizer of $\norm{[\vpsi_x]_i \vpsi_u - \bar{H}^{(i)}\vpsi_x}^2$.
We therefore propose to threshold the least squares error with a user-defined value $\epsilon_H$:
\begin{equation}
  \mathbf{s}_i =
  \begin{cases}
    1, & \text{if} \norm{[\vpsi_x]_i \vpsi_u - \bar{H}^{(i)}\vpsi_x} \leq \epsilon_{H}, \\
    \mathbf{0}, & \text{otherwise}.
  \end{cases}
\end{equation}

The overall procedure for finding $\vs$ and $H$ is summarized in \Cref{alg:block-selection}. The matrix $S$ can be extracted from $\vs$ using the definition of the latter.

\begin{algorithm}[t]
  \caption{Blockwise Identification of Measurement and Selection Matrices}
  \label{alg:block-selection}
  \begin{algorithmic}[1]
    \Require Feature functions $\vpsi_x, \vpsi_u$; error tolerance $\epsilon_{\text{th}}$
    \State Initialize selection vector $\mathbf{s} \gets \mathbf{1}$
    \State Initialize measurement matrix $ H \gets [\;]$ \Comment{empty matrix}
    \State Solve for $\bar H$ using \eqref{eq:hadamard_kron_identity}
    \For{each block $i$ of $\bar H$}
    \If{$\norm{[\vpsi_x]_i \vpsi_u - \bar{H}^{(i)} \vpsi_x} \leq \epsilon_{\text{th}}$}
    \State $H \gets \bmat{H \\ \bar H^{(i)}}$ %\Comment{Stack the block matrix vertically}
    \Else
    \State $[\mathbf{s}]_i \gets 0$
    \EndIf
    \EndFor
  \end{algorithmic}
\end{algorithm}

\subsubsection{Identification of the System Matrix}
Once the selection matrix $S$ is found with the method defined above, we use the bilinear form \eqref{eq:assumption_koopman_bilinear} to find the system matrix $K_x$. In detail, we solve the least squares optimization problem \eqref{eq:edmd} with data matrices of the form discussed in \Cref{sec:trajectory generation} with $\Pin^{(i,j)} \defeq \bmat{\cdots \stack\bigl(\vpsi^{(i,j)}(k), \vpsi^{(i,j)}(k) \otimes \vu^{(i,j)}(k)\bigr) \cdots}$ and $\Pout^{(i,j)}=\bmat{\cdots \vpsi^{(i,j)}_{x}(k+1) \cdots}$, where $k$ is the time step index, $k\in\{0,\ldots,T-1\}$.

\subsection{Control Synthesis via Convex Programming}\label{sec:feedback convergence}
Once the system is identified as described in the previous section, our goal in this section is to design a matrix $K_u$ that stabilizes the closed-loop dynamics $\vpsi_x^+=\tilde{K}\vpsi_x$.
\subsubsection{Semi-Definite Programming (SDP)}
A direct application of \Cref{thm:lyapunov} and \eqref{eq:lti lyapunov simplified} %\Cref{remark:lyapunov simplified}
results in the following Lyapunov condition:
% \begin{equation}
%   \tilde{K}\transpose P \tilde{K}-\lambda P\preceq 0,
% \end{equation}
% which is equivalent to
\begin{equation}\label{eq:lyapunov tilde}
  \lambda P -\tilde{K}\transpose P \tilde{K}\succeq 0.
\end{equation}
Let us fix an arbitrary $P\succ 0$. Applying Schur's complements~\eqref{eq:schur} and the structure of $\tilde{K}$ \eqref{eq:K tilde}, we formulate the following SDP:
%\begin{equation}\label{eq:sdp}
%  \begin{aligned}
%    \min_{\lambda,K_u} &\lambda\\
%    \subjectto&
%                \begin{aligned}
%                  &\bmat{P\inverse & K_{xx}+\sum_{i}^{d_S}K_{xu}^{(i)}K_u H^{(i)}\\
%                  K_{xx}\transpose+\sum_{i}^{d_S}{H^{(i)}}\transpose K_u\transpose {K_{xu}^{(i)}}\transpose & \lambda P}\succeq 0.\\
%                  &\lambda\in[0,1]
%                \end{aligned}
%  \end{aligned}
%\end{equation}

\begin{equation}\label{eq:sdp}
  \begin{aligned}
    \min_{\lambda,K_u}\,& \lambda\\
    \subjectto&
                \begin{aligned}
                  &\bmat{P & P \tilde{K} \\
                  \tilde{K}^\top P & \lambda P}\succeq 0.\\
                  &\lambda\in[0,1]
                \end{aligned}
  \end{aligned}
\end{equation}

\noindent Algorithm \ref{algo:SDP-controller} formalizes this process.
The process is constructive, so no guarantee exists on finding a solution.

\begin{remark}
  \label{remark:convexity_problem}
  In controller design via SDP \cite{boyd1994linear}, a typical step after \eqref{eq:lyapunov tilde} is to pass to the dual of \eqref{eq:lyapunov tilde} and use the Youla change of variable $Y=KQ$, where $Q$ is the dual of $P$; this allows one to obtain an LMI on $Y$ and $Q$. Intuitively, this approach automatically chooses the best Lyapunov function by optimizing over $Q$ as the same time as $Y$; however, in our case, we need to add linear constraints on $\tilde{K}$ to enforce the structure given by \eqref{eq:K tilde}. As far as we can tell, there is no way to formulate this constrained quadratic stabilizability problem as a convex program unless the Lyapunov function is fixed first (which is the approach we take in this paper).
\end{remark}
\begin{algorithm}[b]
  \caption{Semidefinite Program (SDP)-Based Controller Synthesis}
  \label{algo:SDP-controller}
  \begin{algorithmic}[1]
    \State Initialize $K_{u} \gets \mathbf{0}_{d_{\vpsi_u} \times d_u}$, $\lambda \gets \infty$, $P \gets I_{\vpsi_{x}}$
    \While{$\lambda \geq 1$}
    \While{\eqref{eq:sdp} is infeasible} 
    \State Sample $R \sim \mathcal{U}([-1,1]^{n \times n})$
    \State $P \gets R^\top R + \epsilon I_n$ \Comment{Positive definite Lyapunov candidate}
    \State $P \gets A_{\mathrm{dec}}^\top P A_{\mathrm{dec}}$ \Comment{Restrict to state subspace}
    \EndWhile
    \State $(\lambda, K_u) \gets$ Solution of SDP \eqref{eq:sdp}
    \EndWhile
    \State \Return $K_u$
  \end{algorithmic}
\end{algorithm}

%\begin{remark}
%  In practice, if the synthesis problem \eqref{eq:sdp} is not feasible, one can try to increase the size of $\vpsi_x,\vpsi_u$.
%\end{remark}
%\taha{Does it need further explanation?}

To obtain sharper convergence results (i.e., better convergence rates), we add the following assumption (which is common in the literature about control involving Koopman operators \cite{abraham2017model, nozawa2024monte}):
\begin{assumption}
  The observables $\vpsi_x$ include the original state as their first $d_x$ terms.
\end{assumption}
The assumption implies that we can build a \emph{linear decoding operator} $A_{dec}$
\begin{equation}\label{eq:Adec}
  A_{dec} = \begin{bmatrix} I_n \quad \mathbf{0}_{n \times (m-n)} \end{bmatrix}
\end{equation}
such that we can easily extract the state from the vector of observables,
\begin{equation}\label{eq:decoding}
  \vx = A_{dec}\mathbf{\psi}(\vx).
\end{equation}

The operator $A_{dec}$ allows us to enforce convergence only on the original state coordinates (i.e., the first $d_x$ entries of $\vpsi_{\vx}$) rather than the full observable vector. In detail, we sample a random matrix $R \in \mathbb{R}^{d_x\times d_x}$, and define the positive semidefinite matrix $P$ as
\begin{equation}\label{eq:P dec}
  P = A_{dec}\transpose(R\transpose R + \epsilon_P I)A_{dec},
\end{equation}
where $\epsilon_P$ is a user-defined constant.
\begin{proposition}
  If the SDP problem \eqref{eq:sdp} is feasible with $P$ of the form \eqref{eq:P dec} and with optimal value $\lambda^*$, then the first $d_x$ states of the dynamics $\vpsi^+=\tilde{K}\vpsi$ converge with geometric rate $\sqrt{\lambda^*}$.
\end{proposition}
\begin{proof}
  Let $\vx$ denote the first $d_x$ elements of $\vpsi$, and $\norm{x}^2_S=x\transpose S x$ denote the energy norm induced by a positive definite matrix $S=R\transpose R + \epsilon_P I$. Then the linear matrix inequality constraint in \eqref{eq:sdp} implies the Lyapunov condition \eqref{eq:lyapunov tilde}, which in turn implies
  \begin{multline}
    {\norm{\vx}_S^2}^+=
    {\vx^+}\transpose S \vx^+=\\
    {\vpsi^+}\transpose P \vpsi^+=\vpsi\transpose(k)\tilde{K}\transpose P \tilde{K}\vpsi(k)\\
    \leq \lambda^\star \vpsi\transpose P\vpsi = \lambda^\star \vx\transpose S \vx=\lambda^\star \norm{\vx}_S^2.
  \end{multline}
  This implies that $\norm{\vx(k)}_S\leq \sqrt{\lambda^*}^k\norm{\vx(0)}_S$, hence the claim.
\end{proof}
Intuitively, with this choice of $P$, the controller obtained through the SDP \eqref{eq:sdp} does not aim to reduce other auxiliary observables (such as a constant term, $[\vpsi_x]_i = 1$).

As discussed in \Cref{remark:convexity_problem}, the problem is non-convex unless $P$ is fixed. Therefore, starting from identity, we test several random $P$ matrices until feasibility is achieved and the Lyapunov function decreases over time. The procedure is summarized \Cref{algo:SDP-controller}.

% \Cref{example:homing} gives an idea of how our \kcf{} and \Cref{assumption:H} can be used for designing a stable controller in the context of homing with nonlinear distance measurements (see \cite{napoli2023distance} for full details).
% allows us to minimize solve:
% \begin{itemize}
% \item $K_x$ from training data, with $u$ random or generated from a policy parametrized by $K_u$
% \item $K_u$ to either mimic a reference controller, or to obtain properties on $\vpsi_x^+$ (such as convergence \rtronref{stable control design}).n
% \end{itemize}
% \rtron{Give details on the two points above}

% \begin{figure}
%   \includegraphics[width=\linewidth, trim=1cm 7mm 1cm 1.5cm,clip]{figures/koopmanDistanceHoming-paths.png}
%   \caption{Trajectories of the homing controller of \protect\Cref{example:homing} with noisy measurements. Note the stable behavior despite the significant disturbances.}
% \end{figure}

\begin{figure*}[t]
  \centering
  % Row 1: Phase portraits
  \subfloat[Single pendulum phase portrait]{%
    \includegraphics[width=0.45\textwidth]{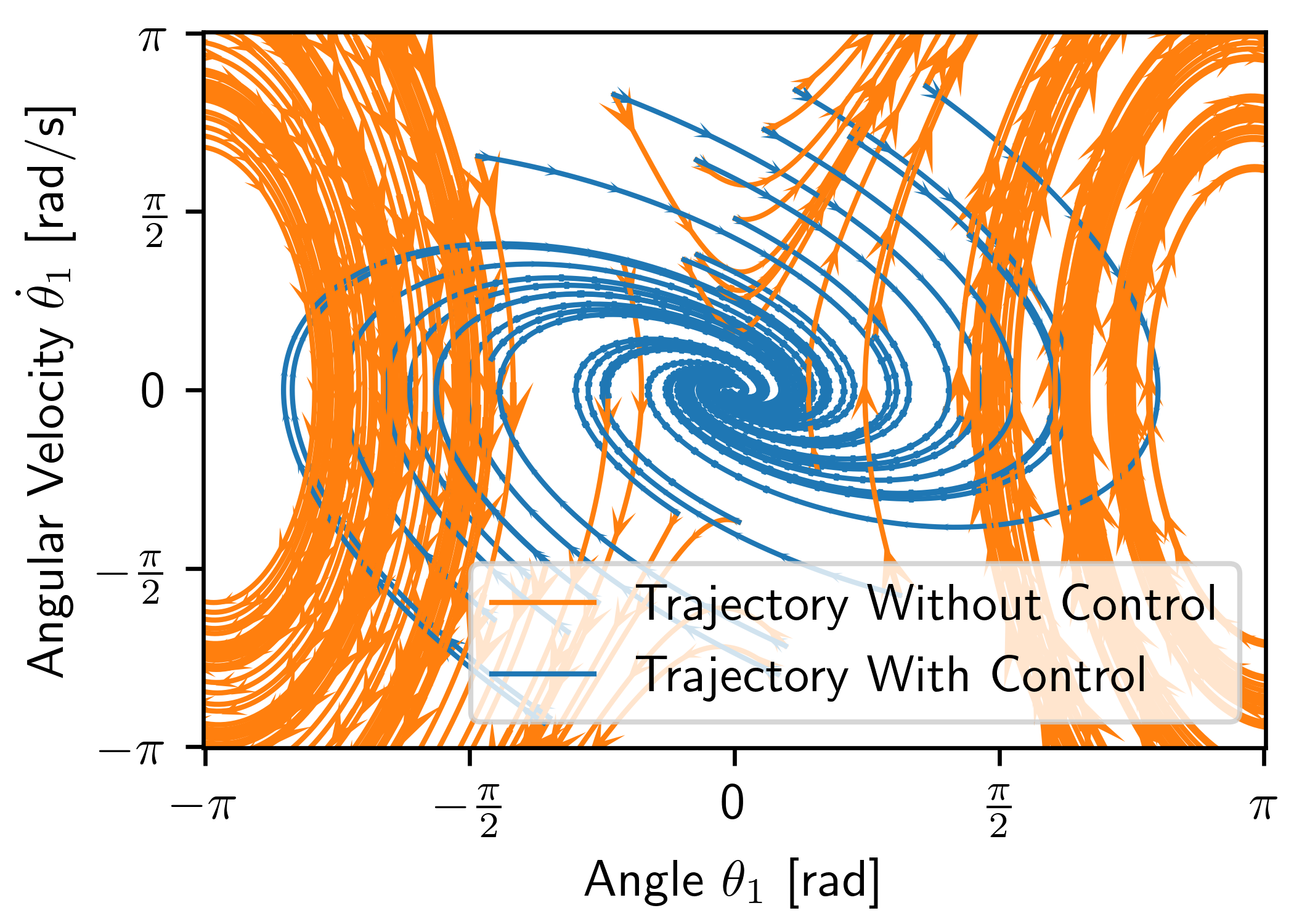}%
    \label{fig:phase-single}%
  }
  \hfill
  \subfloat[Double pendulum phase portrait]{%
    \includegraphics[width=0.45\textwidth]{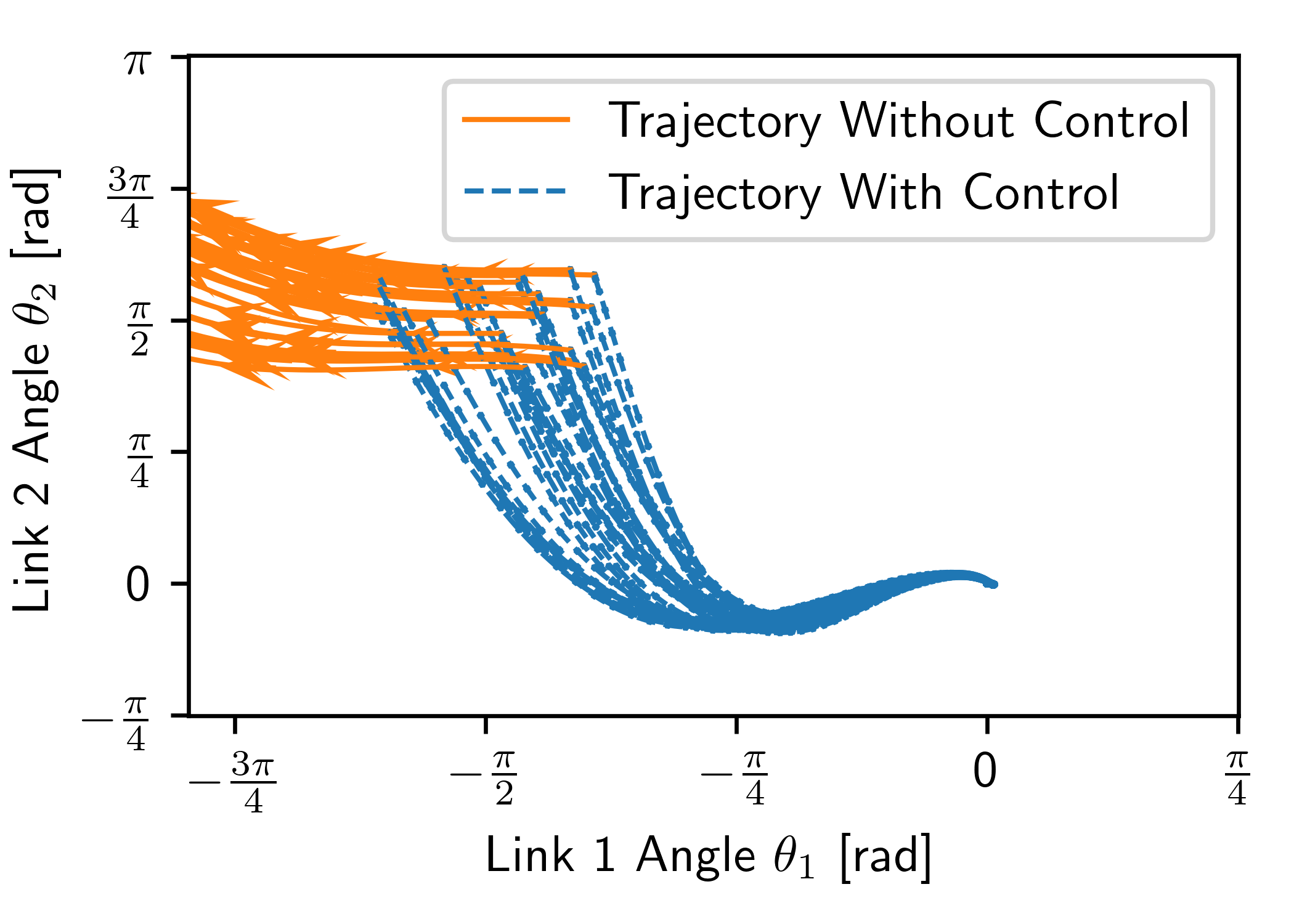}%
    \label{fig:phase-double}%
  }\\[1ex]
  % Row 2: Time evolution
  \subfloat[Single pendulum state response]{%
    \includegraphics[width=0.45\textwidth]{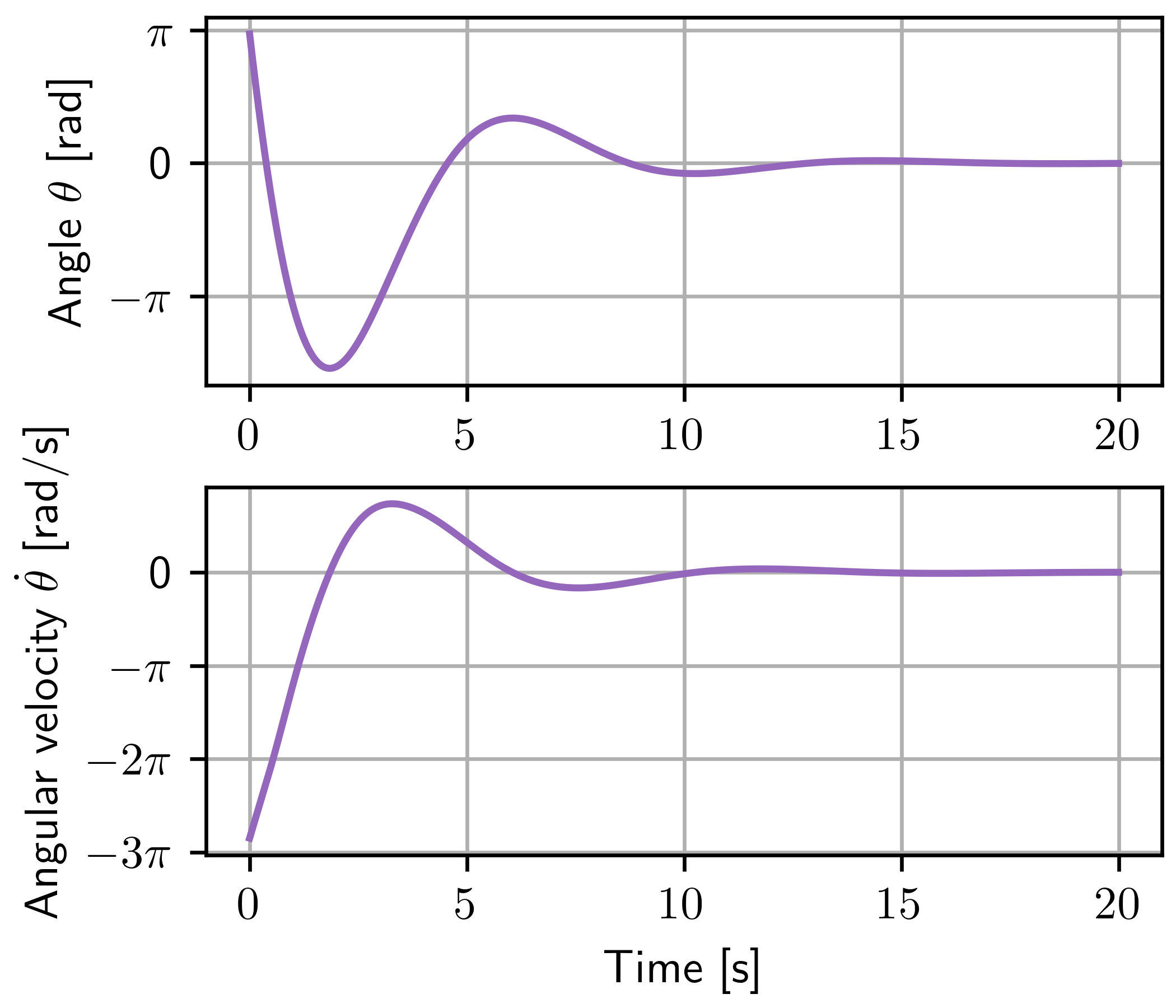}%
    \label{fig:evolution-single}%
  }
  \hfill
  \subfloat[Double pendulum state response]{%
    \includegraphics[width=0.45\textwidth]{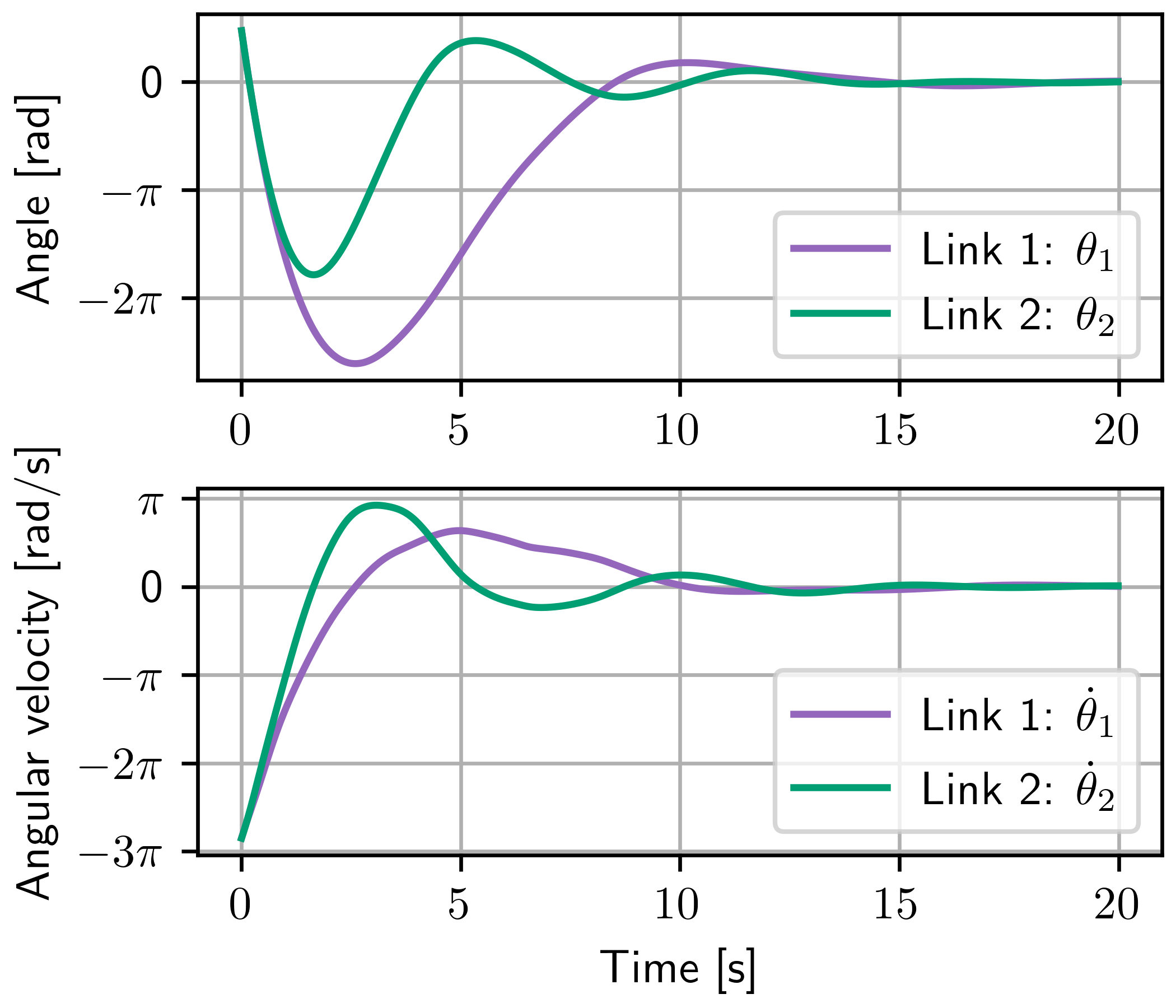}%
    \label{fig:evolution-double}%
  }
  \caption{Uncontrolled and controlled trajectories of the single and double pendulum systems. Thirty trajectories are simulated for each case.
    (a,b) Phase portraits of the uncontrolled and controlled dynamics.
    (c,d) Time evolution of the states under the synthesized controller.}
  \label{fig:controlled-comparison}
  \vspace{-0.3cm}
\end{figure*}

\section{Evaluations of Koopman Control Factorization Framework on Dynamical Systems}
\label{sec:evaluations}

We evaluate our Koopman control factorization framework and associated SDP-based controller on the inverted single and double pendulum with damping, using the dynamics described in \cite{nozawa2024monte}. For the single pendulum, we adopt the parameters point mass $m=1$, link length $L = 1$ and with the damping term $b = 0.3$. Guided by the dynamics, we construct the observable functions including the state variables, a constant term, quadratic terms, and trigonometric features:

\begin{equation}
  \vpsi_u(\vx) = \vpsi_x(\vx) =
  \begin{bmatrix}
    \vx \quad
    1 \quad
    \vx^2 \quad
    \sin(\vx) \quad
    \cos(\vx)
  \end{bmatrix}^\top \in \mathbb{R}^9.
  \label{eq:singlePendulumObservables}
\end{equation}
where the state $\vx$ is defined as $[\theta, \dot\theta]^\top$. 

For the double pendulum, we set the parameters to $m_{1,2} = 1$, and $l_{1,2} = 1$. The state is defined as 
$
    \vx = \begin{bmatrix} \theta_1 & \theta_2 & \dot\theta_1 & \dot\theta_2 \end{bmatrix}^\top.
$
Since the dynamics are governed by trigonometric functions of the link angles and their relative position $\theta_r = \theta_1 - \theta_2$, we incorporate periodic terms at multiple frequencies, such as $\sin(\theta_1)$, $\sin(\theta_r)$, $\sin(2\theta_1)$ and $\sin(2\theta_r)$. These are augmented with the angular velocities, and a constant feature. We further include monomial combinations of these features up to degree 3, (e.g. $\dot\theta^2\sin(2\theta_r)$). To account for the structure of the dynamics, we also include a denominator term that mimics the inversion of the mass matrix.
Guided by iterative feature selection and inspired by the findings of \cite{kaheman2020sindy}, the resulting nonlinear observable map is given by

\begin{equation}
\vpsi_{u}(\vx)={\vpsi}_{x}(\vx) =
\begin{bmatrix}
  \vx \\
  1\\ \frac{1}{3-2\cos\theta_r} 
  \left[
    \begin{array}{c}
      \sin (\theta_1) \\
      \sin (\theta_r) \\
      \sin (\theta_1 - 2\theta_2) \\
      \sin (\theta_r) \cos(\theta_1) \\
      \cos (\theta_r) \sin(\theta_1) \\
      \dot\theta_1^2 \sin (\theta_r) \\
      \dot\theta_2^2 \sin(\theta_r) \\
      \dot\theta_1^2 \sin(2\theta_r) \\
      \dot\theta_2^2 \sin(2\theta_r) \\
      
    \end{array}
  \right]
\end{bmatrix}
\in \mathbb{R}^{14}.
\label{observables_double_pendulum}
\end{equation}
While the selected observables in \eqref{observables_double_pendulum} represent a dynamics-informed and fine-tuned feature map sufficient to demonstrate the proposed \kcf{} framework, systematic feature engineering lies beyond the scope of this work. 

For the single pendulum, the training set consists of 4000 trajectories, generated from initial conditions uniformly sampled on a grid with $\theta \in [-\pi, \pi]$ and $\dot{\theta} \in [-6, 6]$. The control input is constrained to $u \in [-5, 5]$. Each trajectory is simulated for 1 second ($N=100$ steps) using the fourth-order Runge–Kutta (RK4) method.
For the double pendulum, 9000 trajectories are generated from initial conditions sampled on the grid $\theta_{1,2} \in [-\pi, \pi]$ and $\dot{\theta}_{1,2} \in [-6, 6]$, with control inputs $u_{1,2} \in [-5, 5]$. Each trajectory is likewise simulated for 1 second with RK4.
Consequently, the training datasets contain 400{,}000 snapshots for the single pendulum and 900{,}000 snapshots for the double pendulum. The number of datapoints was empirically selected by examining the mean-squared error of the fit in \eqref{eq:koopman affine control}, and by visually comparing the actual trajectories with the Koopman predictions. These datasets are used to identify the Koopman operator matrices as outlined in \Cref{sec:sysID_babbling}. The method described in \Cref{sec:H_sysID} is used to find suitable \emph{selection} and \emph{observation} matrices.

Following the system identification of bilinear Koopman system, we utilize \Cref{algo:SDP-controller} to find a linear control law that acts on nonlinear output functions to stabilize the system. The optimization problem presented in \eqref{eq:sdp} is structured and resolved utilizing the CVXPY package for convex optimization \cite{diamond2016cvxpy}.

The resulting controllers are tested at initial velocities of $[-9, 9]$ for both of the systems, ensuring that the test cases cover a broader range compared to training data, and allowing of the controller’s generalization to unseen scenarios. The tests \footnote{Pendulum stabilization test examples can be found in the \href{https://www.dropbox.com/scl/fo/u34y5mw1kpw1l19ty0cz9/APaiuskzYUaatefs0h-xBLM?rlkey=cdbzg6kg1gdq2sdtuvb5e3xmy&e=1&st=slw7m8l5&dl=0}{video link}.} demonstrate that the controller is capable of stabilizing the system for all the initial conditions, with no steady state error. However, the hand-crafted denominator feature in \eqref{observables_double_pendulum} is necessary for a precise approximation of the original dynamics for the states with relative angle close to zero (states close to singularity). If \eqref{observables_double_pendulum} is implemented without the denominator term, the results have steady-state errors. The hand-crafted denominator feature in \eqref{observables_double_pendulum} is essential for accurately approximating the original dynamics when the angles are close to each other (i.e., near singular configurations). Omitting the denominator term in \eqref{observables_double_pendulum} leads to steady-state errors. Nonetheless, even without this term, the controlled systems remain Lyapunov-stable, as predicted.

Fig.~\ref{fig:phase-single} displays the controlled and uncontrolled trajectories of joint angle and its corresponding velocity, illustrating the performance of the synthesized controller based on our method.  Fig.~\ref{fig:phase-double} depicts the angles ($\theta_1, \theta_2$) of the double pendulum example. To improve visual clarity on the phrase portrait, the initial joint angles are chosen such that $\theta_1$ is sampled within an interval of width $2\pi/9$ centered at $-\pi/2$, and $\theta_2$ within an identically sized interval centered at $+\pi/2$. The initial joint velocities are set to zero, i.e., $\dot{\theta}_1 = \dot{\theta}_2 = 0$. In practice, the controller can stabilize the system from any range of initial conditions between $[-\pi/2,\pi/2]$. Finally, Figs.~\ref{fig:evolution-single} and \ref{fig:evolution-double} show the evolution of the state variables over a 20-second simulation. The initial condition in Fig.~\ref{fig:evolution-single} is set to $\bmat{\tfrac{\pi}{2}, -9}$, while in Fig.~\ref{fig:evolution-double} it is chosen as $[\tfrac{\pi}{2}, \tfrac{\pi}{2}, -9, -9]$. Both of them correspond to a state not included in the training set, which demonstrates that stability is preserved even for previously unseen cases.

\section{Conclusion}
\label{sec:conclusion}
This paper introduced a novel linear output feedback feedback controller synthesis via re-parameterizing bilinear Koopman systems. Although our proposed Koopman-based linear output feedback controller approach yields promising results, it retains many of the same limitations as other work in Koopman theory.
We demonstrate a hand-crafted selection of nonlinear lifting functions for the state and output, but future work could include improved or automated basis function selection for more general mechanical systems.
Since our controller is a proportional controller, the choice of nonlinear lifting functions for output may lead to imperfect approximation for the dynamic system and therefore may cause steady-state errors in control: this issue also remains future work. 
However, stability in the sense of Lypaunov is still guaranteed even if not asymptotic, as long as the underlying dynamics in the entire state space are captured by the Koopman approximation. 

Future directions include a more thorough investigation of Assumptions \ref{assumption:H} and \ref{assumption:K} with respect to their application in hardware systems.
This direction may involve jointly learning the observables with the $S$ and $H$ matrices alongside the control gains. In addition, the structure of \kcf{}  formulation allows for different least squares cost functions and constraints, enabling the emulation of advanced control policies (e.g., MPC) as well as supporting specific objectives such as stability (\Cref{sec:feedback convergence}), imitation of demonstrations, and adaptive modulation.
\bibliographystyle{biblio/IEEEtran}
%\bibliography{biblio/IEEEAbrv,biblio/IEEEConfAbrv,biblio/OtherAbrv,% Do not insert spaces in this command, otherwise it will not work.
\bibliography{
    % Do not insert spaces in this command, otherwise it will not work.
  biblio/IEEEAbrv,biblio/IEEEConfAbrv,biblio/OtherAbrv,
  % biblio/IEEEFull,biblio/IEEEConfFull,biblio/OtherFull,
  biblio/control,biblio/controlAdaptive,biblio/controlCBFs,%
  biblio/koopman,biblio/koopmanControl,biblio/koopmanSoftRobots,%
  biblio/math,biblio/optimization,%
  biblio/tron,%
  %biblio/writing,%
  biblio/education,%
  biblio/formationControl,%
  biblio/websites}

\end{document}